\documentclass[onecolumn]{pasj00}

\begin{document}
\SetRunningHead{T.Okuda}{Super-Eddington Black-Hole Models for SS 433}

\title{Super-Eddington Black-Hole Models for SS 433}

\author{Toru \textsc{Okuda}}
\affil{Hakodate College, Hokkaido University of
Education, 1-2 Hachiman-cho, Hakodate 040-8567}
\email{okuda@cc.hokkyodai.ac.jp}


\maketitle

\begin{abstract}

 We examine  highly super-Eddington black-hole models for SS 433,
based on two-dimensional hydrodynamical calculations coupled with
radiation transport.
The super-Eddington accretion flow with a small viscosity parameter,
 $\alpha = 10^{-3}$, results  in a geometrically and optically 
 thick disk with a large opening angle of $\sim 60^{\circ}$ to the 
 equatorial plane and a very rarefied, hot, and optically thin 
 high-velocity jets region around the disk. 
 The thick accretion flow consists of two different zones: 
 an inner advection-dominated zone and an outer convection-dominated zone.
 The high-velocity region around the disk is divided into 
 two characteristic regions, a very rarefied funnel region along the 
 rotational axis and a moderately rarefied high-velocity region outside of 
 the disk.
 The temperatures of $\sim 10^7$ K and the densities of $\sim 10^{-7}$
  g cm$^{-3}$ in the upper disk vary sharply to $\sim 10^8$ K and
   $10^{-8}$ g cm$^{-3}$, respectively, across the disk boundary 
 between the disk and the high-velocity region.
 The X-ray emission of iron lines would be generated only in a 
 confined region between the funnel wall and the photospheric disk boundary, 
 where flows are accelerated to relativistic velocities of $\sim$ 0.2 $c$  
 due to the dominant radiation-pressure force.
 The results are discussed regarding the  
 collimation angle of the jets, the large mass-outflow rate obserevd 
 in SS 433, and the ADAFs and the CDAFs models.

\end{abstract}

\section{Introduction}
 
 SS 433 is a very unusual and puzzling X-ray source, which exhibits 
 remarkable observational features, such as two oppositely directed 
 relativistic jets moving with a velocity of 0.26 $c$, its expected too-high 
 energy, and a precessing motion of the jets over a period of 162.5 d.
 Although a number of papers concerning observational and theoretical
  data on SS 433 have been published (e.g., \cite{Margo1984}; 
 \cite{Chere1993}), the true nature of SS 433 is  still not clear.
 Before the discovery of SS 433,  Shakura and Sunyaev (1973)
 already discussed the observational appearance of a super-Eddington 
 accretion disk around black holes. 
 Regarding these points, many studies have been made concerning the 
 relativistic jets of SS 433 (\cite{Katz1980}; \cite{Meier1982}; 
 \cite{Lipun1982}) and its collimated ejection (\cite{Sikor1981};
 \cite{Begel1984}).
 Thus, the super-Eddington accretion disks were generally expected to
 possess vortex funnels and radiation-pressure driven jets from
 geometrically thick disks (\cite{Lynde1978}; \cite{Fukue1982}; 
 \cite{Calva1983} ). The radiative acceleration of particles near to
 the disk was also studied by several authors (\cite{Bisno1977}; Icke 1980, 
 1989; \cite{Fukue1996}; \cite{Tajim1998}).
 A two-dimensional hydrodynamical calculation of a super-Eddington 
  accretion disk around a black hole was first examined by Eggum et al.
 (1985, 1988), and was discussed regarding SS 433.
 They showed a relativistic jet formation just outside a conical photosphere 
 of the accretion disk and a collimation angle of $\sim 30^{\circ}$ of the 
 jet, where the dominant radiation-pressure force accelerates the jet to 
 about 1/3 $c$.  However, the jet's mass flux is only 0.4\% of the input 
 accretion rate and is too small for SS 433.

 Kotani et al. (1996) strongly suggested  that SS 433 is 
 a binary system under a highly super-Eddington regime of accretion, where 
 the mass-outflow rate of SS 433 jets is restricted to be 
 $\ge 5\times 10^{-7} M_{\solar}$  yr$^{-1}$, combining spectral 
 X-ray observations of iron emission lines  with hydrodynamical modeling. 
 \citet{Brink2000} have modeled the two-dimensional hydrodynamical outflow
 of the SS 433 jets, while developing their previous model of conically 
 out-flowing jets, and discussed the relationship between the physical
 parameters of SS 433 jets and observationally accessible data.

 \citet{Bland1999} investigated an 
 adiabatic inflow--outflow solution of the accretion disk, where the 
 super-Eddington flow  leads to a powerful wind.
 \citet{King1999} discussed the super-Eddington mass transfer 
  in binary systems and how to accrete in such super-Eddington accretion disks.
 Furthermore, recent theoretical studies and two-dimensional
 simulations of the black holes accretion flows ( \cite{Igumen1999}; 
 \cite{Stone1999}; \cite{Narayan2000}; \cite{Quataert2000}; \cite{Stone2001}; 
 \cite{Abnara2001} ) showed a new development 
 in the advection-dominated accretion flows (ADAFs) and the 
 convection-dominated accretion flows (CDAFs) which may be deeply related to
 the super-Eddington accretion flows.  
 \citet{Okuda2000} have examined the super-Eddington 
 accretion disk model of a neutron star for SS 433 
 by two-dimensional hydrodynamical calculations and discussed 
 the characteristics of the super-Eddington flows.
 In this paper, following the neutron star model, we examine 
 the super-Eddington black-hole models and discuss the results 
 regarding SS 433 and also AVAFs and CDAFs.

\section{ Model Equations}

 A set of relevant equations consists of six partial differential 
 equations for density, momentum, and thermal and radiation energy.
 These equations include the full viscous stress tensor, heating and cooling
 of the gas, and radiation transport.   
 The radiation transport is treated in the gray, flux-limited diffusion
 approximation (\cite{Lever1981}).
 We  use spherical polar coordinates ($r$,$\zeta$,$\varphi$), where $r$
 is the radial distance, $\zeta$ is the polar angle measured from 
 the equatorial plane of the disk, and $\varphi$ is the azimuthal angle.  
 The gas flow is assumed to be axisymmetric with respect to $Z$-axis 
 ($\partial /\partial \varphi=0 $) and the equatorial plane .
 In this coordinate system, the basic
 equations for mass, momentum, gas energy, and radiation energy 
 are written  in the following conservative form (\cite{Kley1989}):

 \begin{equation}
   { \partial\rho\over\partial t} + {\rm div}(\rho\mbox{\boldmath$v$}) =  0,  
 \end{equation}
 \begin{equation}   
  {\partial(\rho v)\over \partial t} +{\rm div}(\rho v \mbox{\boldmath$v$})  = 
   \rho\left[{w^2\over r} + {v_\varphi^2\over r}-{GM_* 
   \over (r-r_{\rm g})^2} \right] -{\partial p\over \partial r}+f_r +{\rm div}
   \mbox{\boldmath$S$}_r
   + {1\over r} S_{rr},  
 \end{equation}
 \begin{equation}  
  {{\partial(\rho rw)}\over \partial t} +{\rm div}(\rho rw\mbox{\boldmath$v$}) 
  = -\rho v_\varphi^2{\rm tan}\zeta-{\partial p\over\partial\zeta}
     +{\rm div}(r\mbox{\boldmath$S$}_\zeta)
    +S_{\varphi\varphi}{\rm tan}\zeta + f_\zeta , 
 \end{equation}
 \begin{equation}    
 {{\partial(\rho r{\rm cos}\zeta v_\varphi)}\over \partial t} 
     +{\rm div}(\rho r{\rm cos}
 \zeta v_\varphi\mbox{\boldmath$v$}) = 
 {\rm div}(r {\rm cos}\zeta \mbox{\boldmath$S$}_\varphi),
\end{equation}
\begin{equation}  
  {{\partial \rho\varepsilon}\over \partial t}+
    {\rm div}(\rho\varepsilon\mbox{\boldmath$v$})
      = -p\;\rm div \mbox{\boldmath$v$} + \Phi - \Lambda, \;\;\;{\rm and}
\end{equation}
\begin{equation}       
  {{\partial E_0}\over \partial t}+ {\rm div}\mbox{\boldmath$F_0$} +
        {\rm div}(\mbox{\boldmath$v$}E_0 +\mbox{\boldmath$v$}\cdot P_0) 
        = \Lambda 
      - \rho{(\kappa +\sigma)\over c}\mbox{\boldmath$v$}\cdot
      \mbox{\boldmath$F_0$} ,
 \end{equation} 
 where $\rho$ is the density, $\mbox{\boldmath$v$}=(v, w, v_\varphi)$ are the
 three velocity components, $G$ is the gravitational constant,
 $M_*$ is the central mass, $p$ is the gas pressure,
 $\varepsilon$ is the specific internal energy of the gas,  $E_0$ is 
 the radiation energy density per unit volume, and $P_0$ is the radiative
  stress tensor. It should be noticed that the subscript "0" denotes
  the value in the comoving frame and that the equations are correct
   to the first order of $\mbox{\boldmath$v$}/c$ (\cite{Kato1998}).
 We adopt the pseudo-Newtonian potential (\cite{Paczy1980})
 in equation (2), where $r_{\rm g}$ is the Schwarzschild radius.
 The force density $\mbox{\boldmath$f$}_{\rm R}=(f_r,f_\zeta)$ exerted 
 by the radiation field is given by
\begin{equation} 
  \mbox{\boldmath$f$}_{\rm R}=\rho\frac{\kappa+\sigma}{c}\mbox{\boldmath$F_0$}, 
\end{equation} 
 where $\kappa$ and $\sigma$ denote the absorption and scattering 
 coefficients and $\mbox{\boldmath$F_0$}$ is the radiative flux 
 in the comoving frame.
 $S=( \mbox{\boldmath$S$}_r,\mbox{\boldmath$S$}_\zeta,\mbox{\boldmath$S$}_
 \varphi)$ denote the viscous stress
 tensor. $\Phi=(S\;\nabla)\mbox{\boldmath$v$}$ is the viscous dissipation 
 rate per unit mass.

 The quantity $\Lambda$ describes the cooling and heating of the gas,
  i.e., the energy exchange between the radiation field and the gas
 due to absorption and emission processes,
 \begin{equation}      
      \Lambda = \rho c \kappa(S_*-E_0), 
\end{equation}
 where $S_*$ is the source function and $c$ is the speed of light. 
 For this source function, we assume local thermal equilibrium $S_*=aT^4$, 
 where $T$ is 
 the gas temperature and $a$ is the radiation constant.
 For the equation of state, the gas pressure is given by the ideal gas law, 
 $p=R_{\rm G}\rho T/\mu$, where $\mu$ is the mean molecular weight 
 and $R_{\rm G}$ is the gas constant. 
  The temperature $T$ is proportional to the specific
 internal energy, $\varepsilon$, by the relation $p=(\gamma-1)\rho\varepsilon
  =R_{\rm G}\rho T/\mu$, where $\gamma$ is the specific heat ratio.  
  To close the system of 
 equations, we use the flux-limited diffusion approximation (\cite{Lever1981}) 
 for the radiative flux:
\begin{equation}
   \mbox{\boldmath$F_0$}= -{\lambda c\over \rho(\kappa+\sigma)}
   {\rm grad}\;E_0, 
\end{equation}
\noindent and
\begin{equation}
   P_0 = E_0 \cdot T_{\rm Edd}, 
\end{equation}
where  $\lambda$ and $T_{\rm Edd}$ are the {\it flux-limiter} and the 
 {\it Eddington Tensor}, respectively, for which we use the approximate
 formulas given in \citet{Kley1989}.
 The formulas fulfill the correct
 limiting conditions in the optically thick diffusion limit and the
 optically thin streaming limit, respectively.
 
 For the kinematic viscosity, $\nu$, we adopt a 
 modified version (\cite{Papal1986}; \cite{Kley1996})
  of the standard $\alpha$- model.
 The modified prescription for $\nu$ is given by
\begin{equation}
  \nu=\alpha\; c_{\rm s}\; {\rm min}\left[H_{\rm p},H\right], 
\end{equation}
 where $\alpha$ is a dimensionless parameter , usually $\alpha$=0.001--1.0,
 $c_{\rm s}$  the local sound speed,  $H$ the disk height, and $H_{\rm p}=p
/\mid 
 {\rm grad}
 \;p\mid$  the pressure scale height  on the equatorial plane.
 
\section{Numerical Methods}

 The set of partial differential equations (1)--(6) is 
 numerically solved by a finite-difference method under initial 
 and boundary conditions.
 The numerical schemes used are basically the same as that described by 
  \citet{Kley1989} and \citet{Okuda1997}. 
 The methods are based on an explicit-implicit finite difference scheme.
  $N_r$ grid points (=150) in the radial direction are spaced logarithmically,
 while $N_{\zeta}$ grid points (=100) in the angular direction are equally 
 spaced, but more refined near the equatorial plane, typically $\Delta \zeta=
 \pi/150$ for $\pi/2 \geq \zeta \geq \pi/6 $ and $\Delta \zeta=
 \pi/300$ for $\pi/6 \geq \zeta \geq 0 $.

\subsection
{  Model Parameters }

  For the central star of SS 433, we assume a Schwarzschild black hole
   with mass  $M_*=10M_{\solar}$
  and examine the structure and dynamics of an accretion disk 
around the black hole and its surrounding atmosphere.
 From an observational constraint of the mass-outflow rate, 
  $\dot M_{\rm loss}$, of $\ge 5\times 10^{-7} M_{\solar}$ 
  yr$^{-1}$ ($ 3 \times 10^{19}$ g s$^{-1}$) in SS 433 (\cite{Kotan1996}), 
 we adopt a mass accretion rate $\dot M_*$ of $8\times 10^{19}$ g s$^{-1}$,
 which corresponds to $ \sim 4\dot M_{\rm E}$, 
 where $\dot M_{\rm E}$ is the Eddington critical accretion rate for the
 black hole, given by
 \begin{equation}
   \dot M_{\rm E} = {48\pi GM_*\over\kappa_{\rm e} c},
\end{equation}
where $\kappa_{\rm e}$ is the electron scattering opacity.
 
\begin{table}
\caption{Model parameter}\label{tab:table1}
\begin{center}
\begin{tabular}{lllllll}
\hline\hline
 Model & ${M_*/\MO}$ & $\dot M_*/\dot M_{\rm E}$ &
 $\dot M_*$ ( g  s$^{-1}$) & $\;\;\alpha$
 & $R_*$ & $R_{\rm max}/R_*$  \\
 \hline
 BH-1  & $\;\;\;\;10$ & $\;\;\;\;4$ &  $\;\;\;8\times 10^{19}$ & 
 $10^{-3}$  & $2r_{\rm g}$ & $ \;\;\;\;220$  \\
 BH-2  & $\;\;\;\;10$ & $\;\;\;\;4$ &  $\;\;\;8\times 10^{19}$ &
 $\; 0.1$   & $2r_{\rm g}$ & $ \;\;\;\;220$  \\
 \hline
 \end{tabular}
\end{center}
\end{table}

 For the  viscosity parameter, $\alpha$, we consider two cases, 
 $10^{-3}$ (BH-1) and 0.1 (BH-2).
 The inner-boundary radius, $R_*$, of the computational domain
 is taken to be $2r_{\rm g}$  and the outer-boundary radius, $R_{\rm max}$,
  is selected so that the radiation pressure is comparable to the gas 
 pressure at the outer boundary, where the Shakura--Sunyaev 
 instability never occurs.
 The model parameters used are listed in table \ref{tab:table1}.
 
 \subsection
 {  Boundary and Initial Conditions}

   Physical variables at the inner boundary, except for the velocities,
    are given by extrapolation of the variables near the boundary. 
  However, for the velocities,
   we impose limited conditions that the radial velocities are always 
   negative and the angular velocities are zero.
  If the radial velocity by the extrapolation is positive,
  it is set to be zero; that is, outflow at the inner boundary is 
  prohibited.    
 On the rotational axis and the equatorial plane, 
 the meridional tangential velocity $w$ 
 is zero and all scalar variables must be symmetric relative to these axes.
 
 The outer boundary at $r=R_{\rm max}$ is divided into two parts. 
 One is the disk boundary through which matter is entering from 
 the outer disk.
 At the outer-disk boundary we assume a continuous inflow of matter 
 with a constant accretion rate, $\dot M_*$.   
 The other is the outer boundary region above the
 accretion disk. We impose free-floating conditions on this outer boundary 
 (i.e., all gradients vanish) and allow for outflow of matter, whereas 
 any inflow is prohibited here. 
 We also assume the outer boundary region above the disk to be 
 in the optically-thin limit, 
  $\vert \mbox{\boldmath$F_0$} \vert \rightarrow c E_0$.
 This imposes a boundary condition on the radiation energy density, $E_0$. 
  The initial configuration  consists of a cold, 
 dense, and optically thick disk and a hot, rarefied, and optically
 thin atmosphere around the disk. 
 The initial disk is approximated by the Shakura-Sunyaev's standard
 model.
 The initial hot rarefied atmosphere around the disk is constructed to be
 approximately in hydrostatic equilibrium.

\section{Results}

\subsection{Model BH-1 with $\alpha=10^{-3}$}
 The initial disk thickness, $H/r$, based on the standard disk model, 
 is $\sim 1$ at the inner region
 and  $\sim 0.1$ at the outer boundary.
 The ratio $\beta$ of the gas pressure to the total pressure 
 at the initial outer disk boundary is  $\sim 0.27$ . 
 We performed a time evolution of the disk until $t$ = $4 \times
 10^3 P_{\rm d} $ for model BH-1, where $P_{\rm d}$ is the Keplerian
 orbital period at the inner boundary. 
 
  \begin{figure}
 \begin{center}
 \FigureFile(85mm,75mm){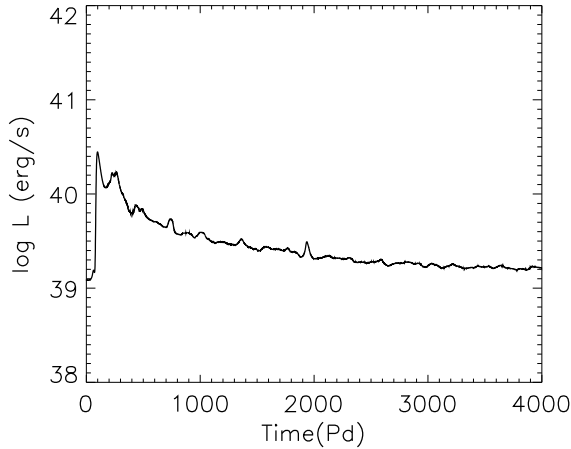}
 \end{center}
 \caption {Time evolution of the luminosity, $L$, for model BH-1, where
  $P_{\rm d}$ is the Keplerian orbital period at the inner boundary
  $2r_{\rm g}$.}\label{fig:fig1}
 \end{figure}
 
 Figure \ref{fig:fig1} shows  the time evolution of the total luminosity, 
 $L$, for model BH-1.
 After an initial sharp rise of the luminosity, the luminosity curve
 descends gradually toward a steady state value.
 The luminosity $L$ at the final phase 
 is $ 1.6 \times 10^{39}$ erg s$^{-1}$ and $L/L_{\rm E}$ is $\sim 1$, 
 which is far smaller than $\dot M_*/\dot M_{\rm E}(= 4)$,
 where $L_{\rm E}$ is the Eddington luminosity.

  At the initial stage of the evolution,  matter in the inner region of
 the disk is ejected  strongly outward.
 Some of the ejected gas hit on the rotational axis and others
 propagate outward.
 The gas hitting on the axis leads to a high-temperature region along
 the axis and anisotropic radiation fields, in which equ-contour
 lines of the radiation energy density are concentric.
 This results in an outward
 radiation-pressure force, which dominates the gravitational force, and
  a high-velocity jet region is formed along the axis.   
 At the beginning of the evolution, the high-velocity region 
 is confined to a narrow region along the rotational axis, but 
 gradually spreads outward from the axis. The spreading-jet gas interacts 
 with the surrounding medium and finally settles down to a quasi-steady state.
 On the other hand, the initially optically thin atmosphere above the initial
 disk is occupied by the dense gas ejected from the inner disk with 
 increasing time, and becomes optically thick at the final phase, 
 forming a cone-like shape with a large opening angle of $\sim 60^{\circ}$ 
 to the equatorial plane.  
 We regard this optically thick and dense region as an accretion disk. 
 The boundary  between the high-velocity region and the disk
 is characterized by  sharp gradients of density and temperature. 
 
 \begin{figure}
 \begin{center}
 \FigureFile(160mm,180mm){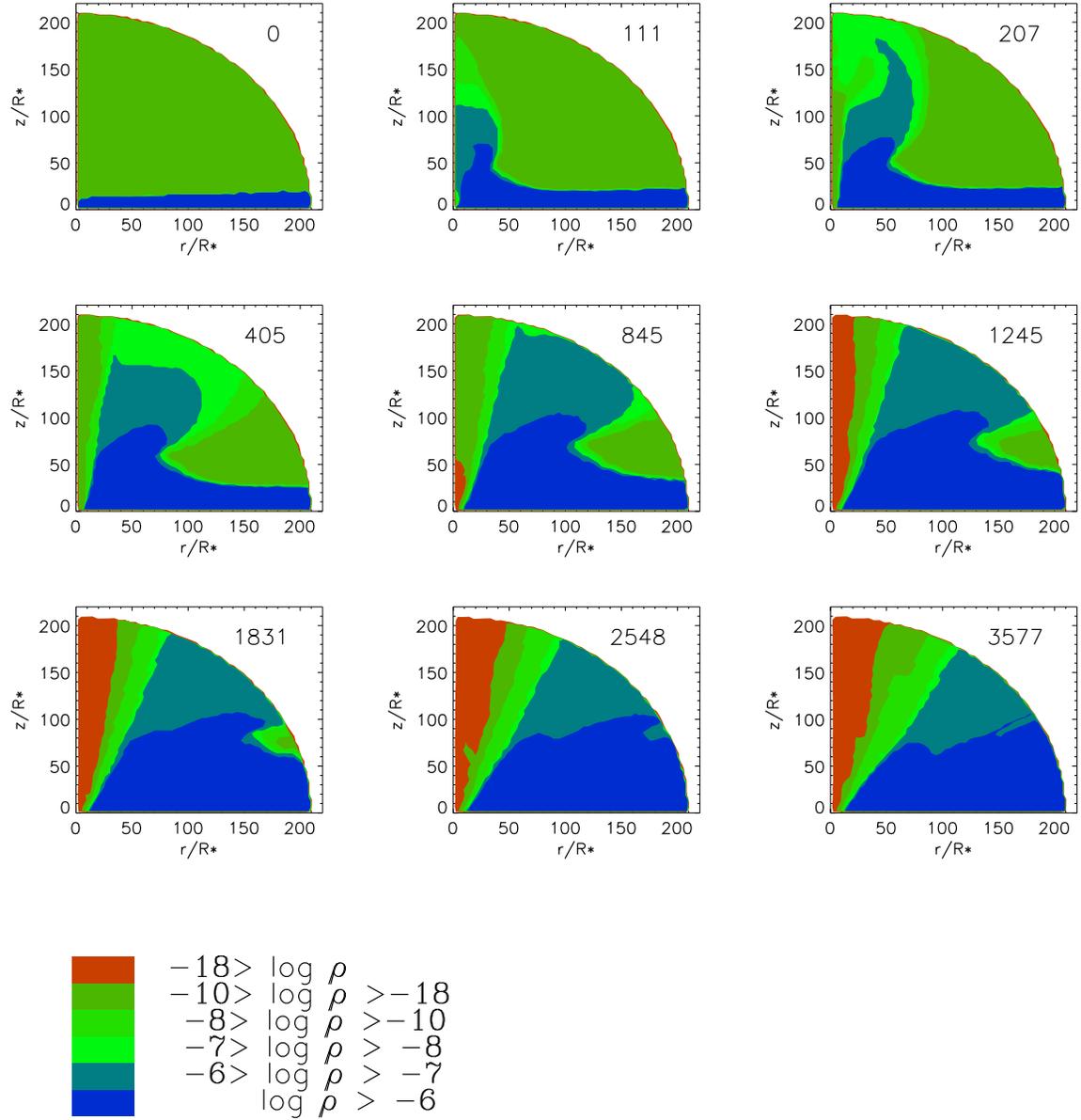}
 \end{center}
  \caption{Snapshots of the density evolution for 
  model BH-1, where the right-hand number in each photo shows the 
  evolutionary time in units of $P_{\rm d}$; also, a color legend 
  for ${\rm log}\; \rho\; ({\rm g}\;  {\rm cm}^{-3})$ is shown. 
  The first snapshot denotes the initial density configurations, which
  consist of the Shakura--Sunyaev disk and the surrounding rarefied 
  atmosphere. 
  The green and orange parts in the later snaps show the hot, rarefied, and 
  optically thin high-velocity jets region,    
  while the blue and dark blue parts denote the cold, dense, and 
  optically thick disk region.
  The flow velocities at the final phase are $\sim 0.1$--0.2 $c$ 
  in the deep-green part and $\sim 0.2$--0.4 $c$ in the light-green and 
  orange parts.
   The observed X-ray emission of iron lines in SS 433 would be formed in 
   the deep-green high-velocity region  just outside the disk boundary.
 } \label{fig:fig2}
 \end{figure}

  In order to understand well the evolution of the disk and the high-velocity
 region, we animated the density and temperature evolutions.
 Figure \ref{fig:fig2} denotes nine snapshots of the density evolution, 
 where the right-hand number in each snap shows 
 the evolutionary time in units of $P_{\rm d}$; also, a color legend for 
  ${\rm log}\; \rho\; ({\rm g}\;  {\rm cm}^{-3})$ is shown.  
  The green and orange parts in the later snaps express the rarefied, 
  hot, and optically thin high-velocity jets region, 
  while the blue and dark blue parts denote the optically thick and 
   dense disk region. 
 In the high-velocity region, we distinguish the orange region
 (which we call the funnel region) from the green region, 
 because  the densities in the orange region decrease sharply towards the
  rotational axis and 
 are as very low as $10^{-18}$--$10^{-26} \;{\rm g}\;  {\rm cm}^{-3}$,
 compared with $10^{-8}$--$10^{-18}\; {\rm g}\;  {\rm cm}^{-3}$ in 
 the green region.  
 Therefore,  we have three characteristic regions here: 
 (A) the disk region (blue), (B) the  high-velocity region (green),
 and (C) the very rarefied funnel region (orange).
 As discussed later,  the centrifugal barrier and the funnel wall lie 
 near the boundaries, between (A) and (B), and between (B) and (C) 
 at the final phase, respectively. 
  
   \begin{figure}
 \begin{center}
 \FigureFile(85mm,155mm){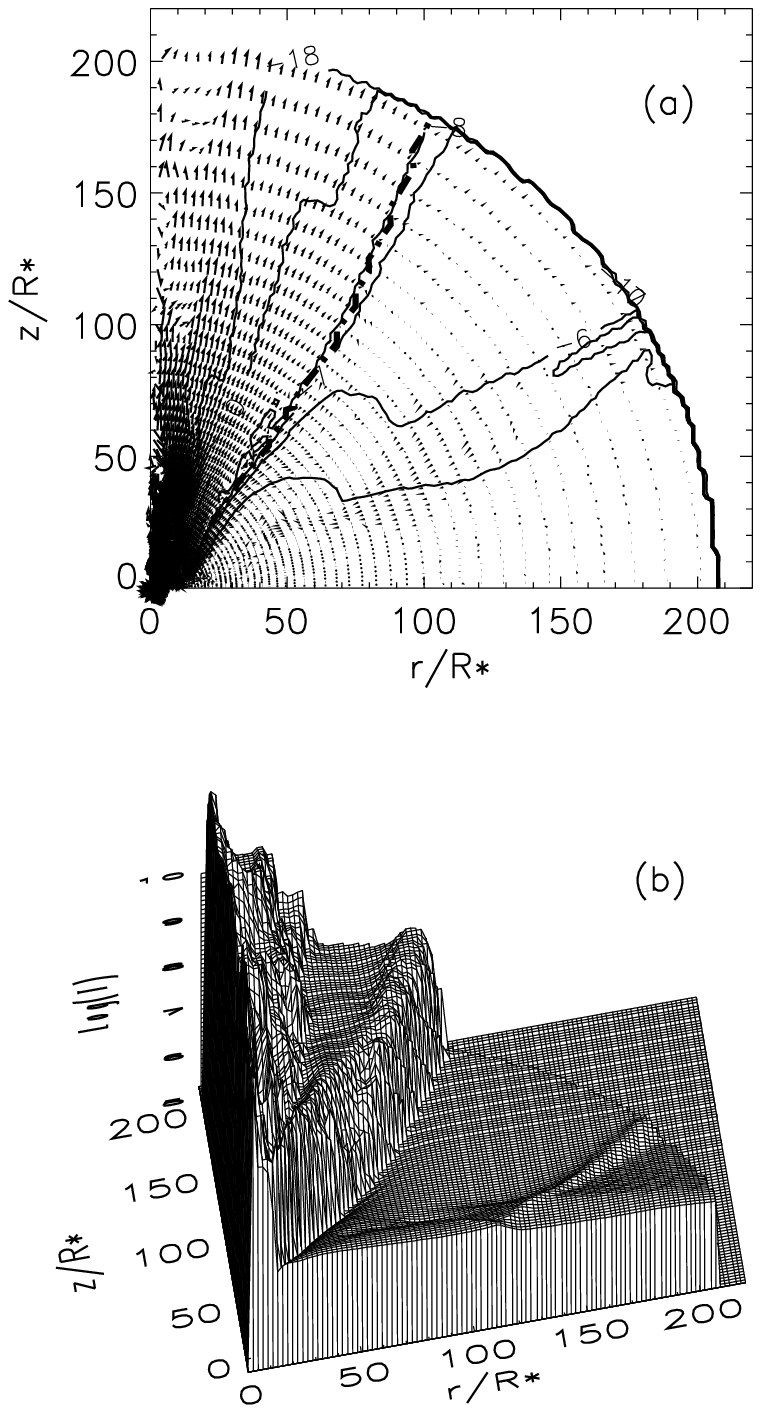}
 \end{center}
 \caption{Velocity vectors and contours of the density, $\rho$
 (g s$^{-3}$), in logarithmic scale (a) and bird's-eye views
 of the gas temperature $T$(K) (b), on the meridional plane at the
 evolutionary time $t=3577 P_{\rm d}$ for model BH-1.
 The density contours are denoted by the labels log $\rho = -5, -6, 
 -7, -8, -10,$ and $-18$, and the velocity vectors show the maximum
 velocity  0.4 $c$ at $\zeta \sim 80^{\circ}$ and $r/R_* \sim 200$.
 The thick dashed  and the thick dot-dashed lines in (a) show 
 the funnel wall and the disk boundary between the disk and the 
 high-velocity region, respectively.}\label{fig:fig3}
 
 \end{figure}
 
 Figure \ref{fig:fig3} shows the density contours with velocity vectors (a) and
 the temperature contours (b) on the meridional plane at $t= 3577 P_{\rm d}$.
 In figure 3a, the density contours are denoted by the 
 labels log $\rho = -5,-6,-7,-8,-10,$ and $-18$. 
  The velocity vectors show the maximum velocity,  0.4 $c$, at 
  $ \zeta \sim 80^{\circ}$ and $r/R_* \sim 200$.
  The thick dashed  and the thick dot-dashed lines in (a) show 
 the funnel wall and the disk boundary between the disk and the 
 high-velocity region, respectively.
 Here, we approximately define the disk boundary as an interface through 
 which a temperature of $\sim 10^7$ K in the disk jumps to $\sim 10^8$ K.
  The disk boundary lies 
  roughly between the density  labels  $-7$ and $-8$.
  In the funnel region (C) ( $ \zeta \gtrsim 80^{\circ}$), the flow 
  velocities are very large, but rather chaotic near the rotational axis.
  In the high-velocity region (B) ( $ 80^{\circ} \gtrsim \zeta \gtrsim 60^
  {\circ}$), the flows are relativistic to be $\sim 0.1$--0.2 $c$, and 
  the gas streams run radially. 
  Figure 3b shows a bird's-eye view of the temperature, where the disk 
  boundary is clearly recognized by a sharp wall with high temperatures. 
  The temperatures range from  $2 \times 10^6$ to  $ \sim 10^7$ K in the
  disk region, jump to $\sim 10^8$ K just across the disk boundary, 
  and again distribute gradually between $\sim 10^8$ and $10^9$ K 
  in the high-velocity region.  In the funnel region near the rotational axis,
 the temperatures  are as very high as $\sim
 10^9$--$10^{11}$ K. On the  mid-plane, the densities and the 
 temperatures  are not largely different from the initial ones, except 
 for the innermost region of $ r/R_* \lesssim 10$, where the flow is  more 
 rarefied and hotter than the initial flow.
    
  \begin{figure}
 \begin{center}
 \FigureFile(85mm,155mm){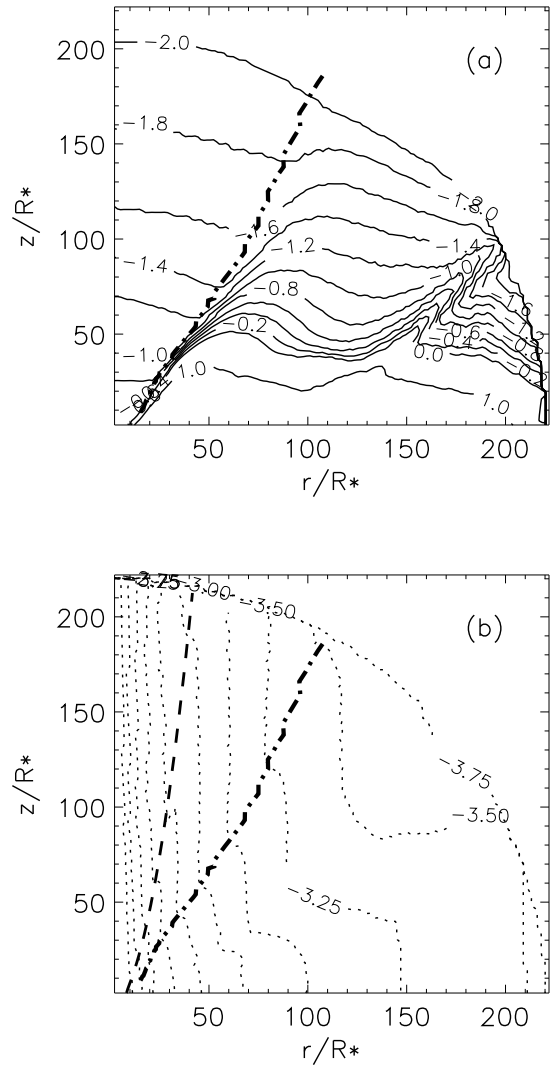}
 \end{center}
  \caption{Contours of the logarithmic radiation energy density log 
  $E_{0}$ in an arbitrary unit (a) and the logarithmic angular-velocity 
  log $\Omega$ normalized by the Keplerian angular velocity at the 
  inner boundary with intervals of 0.25 (b) on the meridional plane 
 at $t = 3577 P_{\rm d}$ for model BH-1, where the thick dashed and the 
 thick dot-dashed lines denote the funnel wall and the disk boundary 
 between the disk and the high-velocity jets region, respectively.
 }\label{fig:fig4}
 \end{figure}  
  
 Figure \ref{fig:fig4} shows the contours of radiation energy 
 density, $E_{0}$, in an
 arbitrary unit (a) and angular-velocity $\Omega$  (b)
 on the meridional plane at $t= 3577 P_{\rm d}$, where the thick 
 dot-dashed line denotes the disk boundary. 
  In figure 4a, $E_0$ is almost independent 
 on $\zeta$ in the optically thin high-velocity region and weakly
 dependent on $\zeta$ in the upper disk region, and so the radiation-pressure 
 gradient forces act only radially in the high-velocity region (B).
 In figure 4b, the angular velocities, $\Omega$, in unit of the Keplerian 
 angular velocity at the inner boundary are denoted by the contours of 
 log $\Omega$.
 The angular velocities show  two kinds of characteristic distributions.
 Near to the equatorial plane, the angular velocities are 
 nearly Keplerian. However, in the upper disk and the high-velocity region, 
 the equi-contours of $\Omega$ are parallel to the $Z$-axis, and 
 $\Omega$ behaves approximately as $K/x^{2.1}$, where $x$ is the 
 distance in units of $2r_{\rm g}$ from the $Z$-axis and $K$ is a constant,
 although the constant $K$ seems to change slightly on 
 the disk boundary and the funnel wall.
 As a result, the specific angular momentum  $\lambda_{\rm a} 
 ( = x^2 \Omega)$ is approximately conserved in these regions. 
 From the numerical data, we find $\Omega = 5.6/x^{2.1}$ with our 
 non-dimensional units in the high-velocity region. 
 The funnel wall, a barrier where 
 the effective potential due to the gravitational potential and  
 the centrifugal one vanishes, is described by a surface $(x_{\rm f},
  z_{\rm f})$, 
 \begin{equation}
       (\Phi)_{\rm eff} =  { - 1 \over (r_{\rm f}- 1/2) }+ 
                 {{\lambda_{\rm a}}^2 \over {2 x_{\rm f}^3 }}= 0,
 \end{equation}
  where $r_{\rm f} = ({x_{\rm f}}^2 + {z_{\rm f}}^2)^{1/2}$ is given 
  in units of $2r_{\rm g}$ (\cite{Molte1996}).
 From equation (13) together with the above $\Omega$, 
 the funnel wall is shown by the thick dashed line in figure 4b.
 The funnel region (C) and the high-velocity region (B) are roughly 
  separated by the funnel wall. 
 
 The acceleration  for the relativistic flows
 depends on the distribution of 
  the radiation pressure, $P_{\rm r}(= f_{\rm E} E_0)$, where $f_{\rm E}$
 is the Eddington factor.
 The flux limiter, $\lambda$, almost equals to 1/3 in the optically thick 
 disk, while in the optically thin high-velocity region, $\lambda$ is 
 very small and its spatial variation is almost identical to that of the 
 density, $\rho$. 
 In regions (B) and (C),  the radiation temperatures, 
 $T_{\rm r} = (E_0/a)^{1/4}$, depend almost only on $r$ and are small
 compared with the gas temperature, $T$;  
  also, the gas temperatures  in the region (C) are 
 much higher than that in the region (B). However, its too high temperature
 ($\sim 10^9$--$10^{11} {\rm K}$) in region (C) may be unreliable
 because we did not take account of other physical processes, such as Compton 
 processes and pair production--annihilation,  which may be important 
 at such high temperatures. On the other hand, we consider that the densities 
 would not be drastically altered by the processes.  
    
 The radiation pressures are very dominant everywhere, especially in the 
 inner region of the disk. In the disk 
 region (A), the centrifugal forces balance the gravitational forces near to
 the disk midplane, whereas, at the upper disk region, the balances due to
 the centrifugal, the gravitational, and the radiation-pressure gradient 
 forces are maintained. In regions (B) and (C), 
 the dominant radial radiation-pressure forces are one order of 
 magnitude larger than the gravitational forces because the densities in 
 these regions are very low.
 As a result, the radiation-pressure gradient forces accelerate  
 the flows radially to relativistic velocities. 
 The azimuthal velocities in the high-velocity region  are in the range 
 of 0.1--0.2 $c$, and the high-velocity gas is blown off spirally and 
 relativistically.
 At the disk boundary between regions (A) and (B), the gravitational 
 force balances roughly the radial component of centrifugal force, i.e., 
 the boundary region corresponds to what is called a centrifugal barrier. 
 
 \begin{figure}
 \begin{center}
 \FigureFile(85mm,125mm){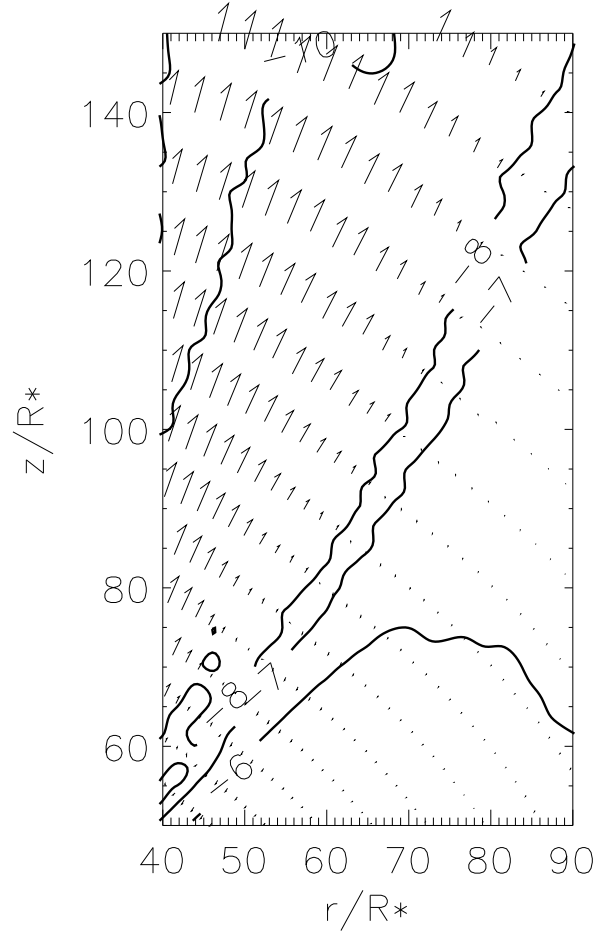}
 \end{center}
  \caption{Magnified velocity fields and density contours near the upper
   disk boundary at $t = 3577 P_{\rm d}$ for model BH-1, where the maximum 
   velocity is 0.22 $c$ and the contours of 
   log $\rho$ (g s$^{-3}$) is shown by lines with labels $-6, -7, -8,$
   and $-10$. This high-velocity region above the disk boundary is a typical
    X-ray emitting region of iron lines with velocities of $\sim$ 0.2 $c$.
   }\label{fig:fig5}
  \end{figure}  
  
 Figure \ref{fig:fig5} shows magnified velocity fields and the 
 density contours 
 near the upper disk boundary at $t = 3577 P_{\rm d}$, where the maximum 
 velocity is $ 0.22 c$ and the contours of log $\rho$ (g s$^{-3}$) is shown 
 by lines with labels $-6, -7, -8,$ and $-10$. The disk boundary between 
 regions (A) and (B) is between the density labels  $-7$
  and $-8$.
 As mentioned later, this high-velocity region above the disk boundary
 would correspond to a typical X-ray emitting region of iron lines 
 with velocities of $\sim$ 0.2 $c$.
 
 Figure 6 shows the flow features in the inner disk and the surrounding
 high-velocity region, where the velocities are indicated by unit vectors
 and the thick dot-dashed line shows the disk boundary.
  One of the remarkable features of luminous accretion disks is convective 
  phenomena in the inner region of the disk (Eggum et al. 1985, 1988;
  \cite{Milso1997}; \cite{Fujita1998}). 
  The convective motions are clearly found in this figure, and there appear
   more than a dozen of convective cells.

   \begin{figure}
 \begin{center}
 \FigureFile(120mm,50mm){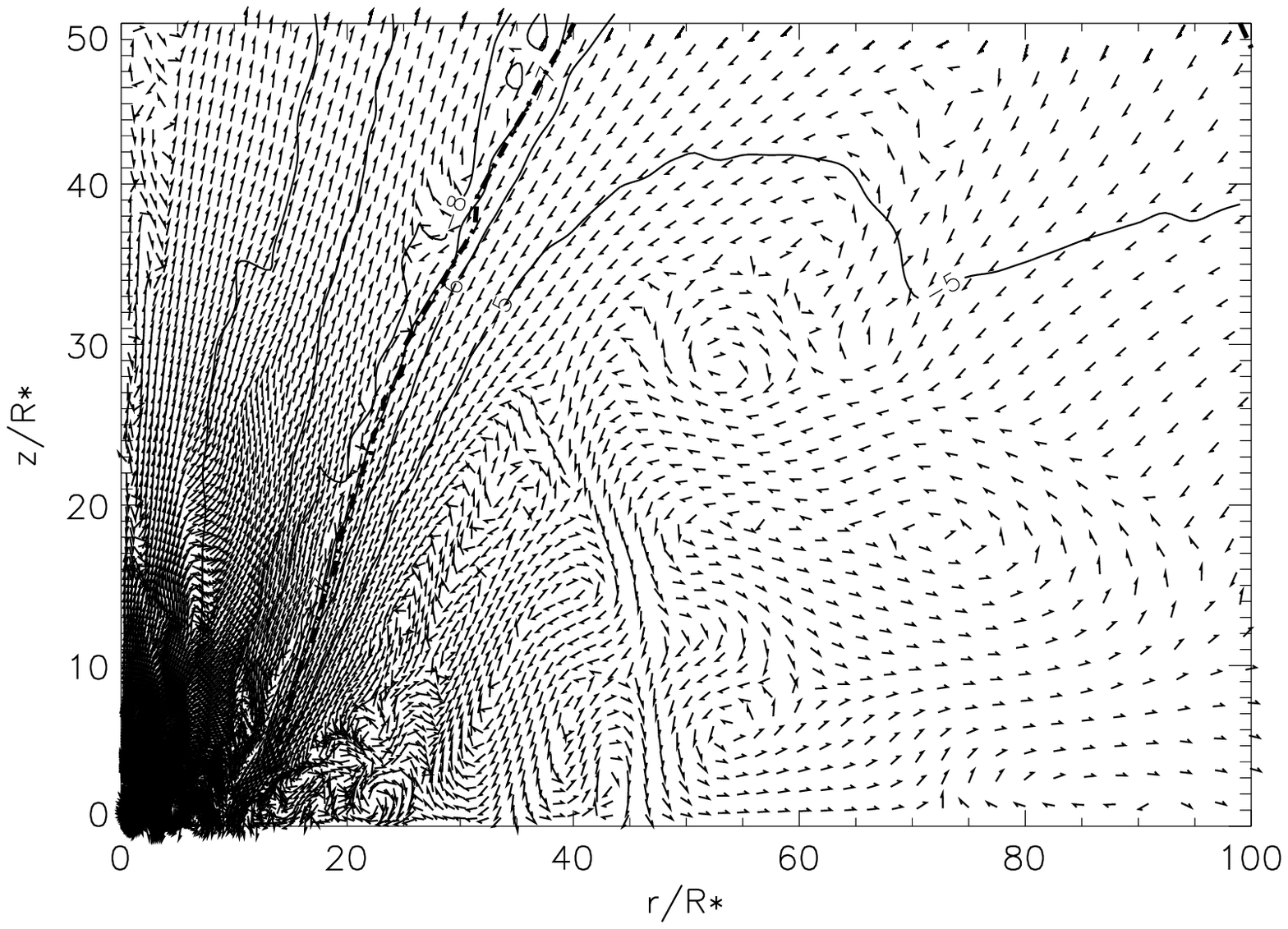}
 \end{center}
  \caption {Unit-velocity vectors and density contours of log $\rho$ =
  $-5, -6, -7, -8, -10,$ and $-18$ in the inner disk and the surrounding 
  high-velocity jets region at $t = 3577 P_{\rm d}$ for model BH-1,
   where the dot-dashed line denotes the disk boundary bewteen the disk 
   and the high-velocity jets region.  
   A transition between the inner advection-dominated zone and the outer
 convection-dominated zone occurs at $r \sim 10R_*(= 20r_{\rm g})$.
 }\label{fig:fig6}
 \end{figure}
 
  Previously it has been shown that a luminous accretion disk with 
  $\dot M_* \sim \dot M_{\rm E}$ is unstable against convection, 
  and that this is induced by a large negative gradient of entropy vartical to 
  the equatorial plane (\cite{Bisno1977}; \cite{Fujita1998}). 
  The accretion flow is severely disrupted by the convective eddies.
  Recent theoretical studies and two-dimensional hydrodynamical simulations of 
  radiatively inefficient black hole accretion flows 
   (\cite{Stone1999}; \cite{Igumen1999}; \cite{Narayan2000}; 
   \cite{Quataert2000}) showed that accretion flows with low viscosities 
   are generally convection-dominated flows (ADAFs and CDAFs)
   and have characteristic self-similar solutions of the disk variables
   which are described by power law profiles with radius. 
   This was also confirmed in  recent magnetohydrodynamical simulations of 
   the accretion flows around black holes (\cite{Stone2001}).

 \begin{figure}
 \begin{center}
 \FigureFile(140mm,200mm){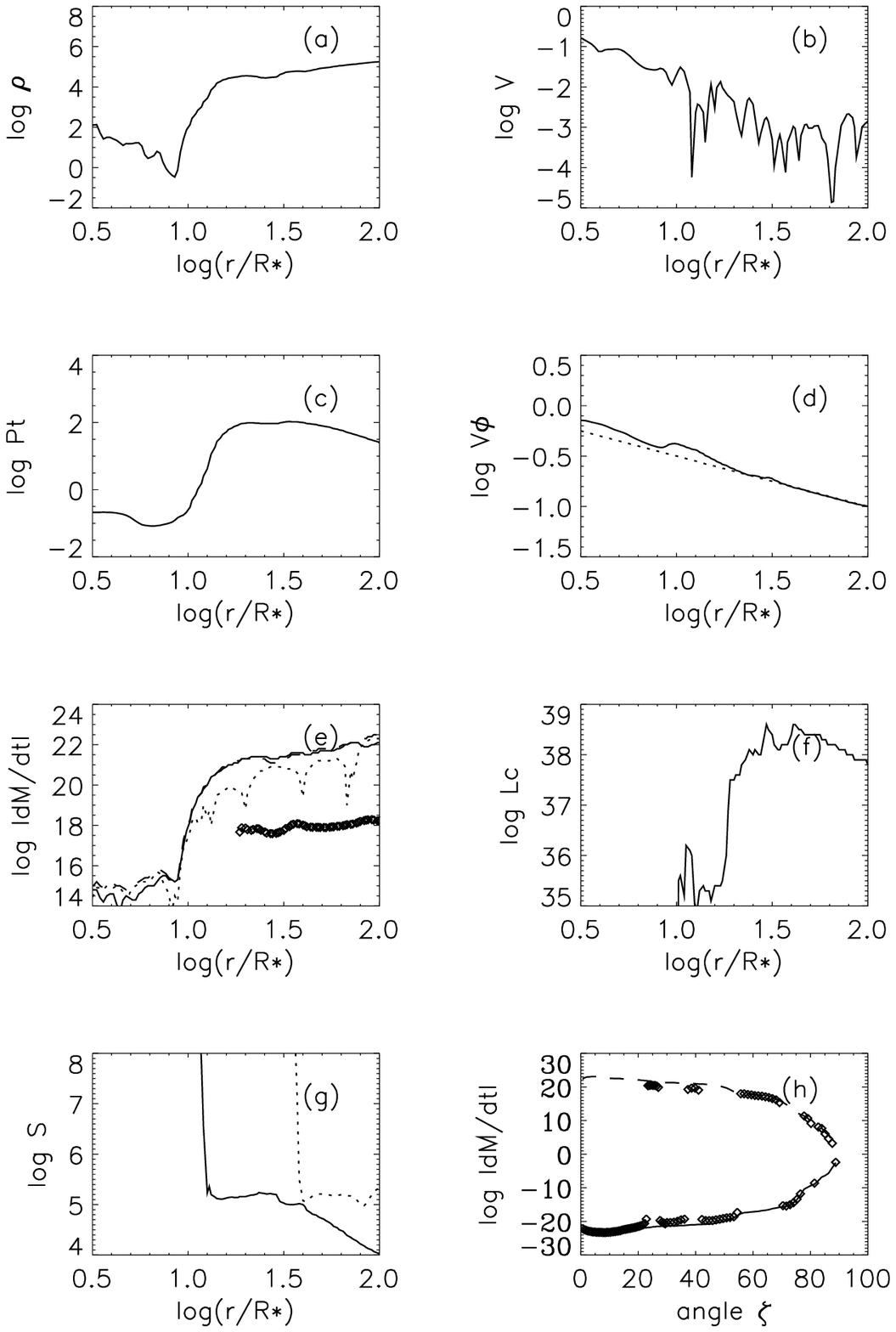}
 \end{center}
  \caption {Radial and angular profiles of the time-averaged flow variables 
  near to the equatorial plane  for model BH-1: 
  (a) density $\rho$ , 
  (b) radial velocity $v$,  
  (c) total pressure $P_{\rm t}$, 
  (d) rotational velocity $v_{\rm \varphi}$, 
    (e) four radial mass-flux rates $\dot M(r)$, 
   (f) convective luminosity  $L_{\rm c}$,  
  (g) entropy $S$,  
   and (h) angular profiles of three mass-flux rates $\dot M(\zeta)$ 
   (see text in detail).

 }\label{fig:fig7}
 \end{figure} 
 
   Model BH-1  belongs to a typical ADAFs or a CDAFs case because of
   its small value of the viscosity parameter, $\alpha$. 
  Figure 7 denotes the radial and angular profiles of the time-averaged 
  flow variables near the equatorial plane  for model BH-1:
  (a) density, $\rho$, in units of $\rho_0(=10^{-8}$ g cm$^{-3})$;
  (b) radial velocity, $v$, in units of Keplerian velocity $v_{\rm K_*}$
  at $r = R_*$;
  (c) total pressure, $P_{\rm t}$, in units of $\rho_0\; v_{\rm K_*}^2$;
  (d) rotational velocity, $v_{\rm \varphi}$ (solid line), in units of
  $v_{\rm K_*}$ and the Keplerian one (dotted line); (e) radial mass-inflow 
  rate, $\dot M_{\rm in}(r)$ (g s$^{-1}$) (solid line),
  mass-outflow rate, $\dot M_{\rm out}(r)$ (dashed line),
   net mass accretion rate, $\dot M_{\rm a}(r)(=\dot M_{\rm in}+\dot M_{\rm
  out})$ (dotted line), in the whole region, and mass-outflow rate 
  $\dot M_{\rm jet}(r)$ ($\diamond$) in the jet region; (f) convective 
  luminosity,  $L_{\rm c} = \int F_{\rm c}(r)dS$ (erg s$^{-1})$;
  (g) entropy, $S$, at $\zeta \sim 0^{\circ}$ (solid line) and $42^{\circ}$
  (dotted line) in an arbitrary unit; and (h) angular profiles of
  mass-inflow rate, $\dot M_{\rm in}(\zeta)$ (g s$^{-1}$) (solid line), 
  mass-outflow rate, $\dot M_{\rm out}(\zeta)$ (dashed line), 
  and net mass-flow rate, $\dot M_{\rm p}(=\dot M_{\rm in}+\dot 
  M_{\rm out}) (\diamond)$, where $F_{\rm c}$ in (f) is the convective 
   energy flux given by \citet{Narayan2000} and its integral is taken within 
   the disk region, and the ordinate scale in (h) shows ${\it Sign}(\dot M$) 
   log $|\dot M|$.  
   The time-averaged and angle-integrated mass inflow and
    outflow rates ($\dot M_{\rm in}$ and $\dot M_{\rm out}$, respectively) 
    in figure 7e are defined as
    
 \begin{equation}
       \dot M_{\rm in}(r) = 4\pi r^2 \int_{0}^{\pi/2} {\rho\; {\rm min}\;(v,0) 
         {\rm cos}\; \zeta} d \zeta, 
 \end{equation}
 \noindent and
 \begin{equation}
       \dot M_{\rm out}(r) = 4\pi r^2 \int_{0}^{\pi/2} {\rho\; {\rm max}\;(v,0)          {\rm cos}\; \zeta} d \zeta.
 \end{equation}
 For these rates, the net mass accretion rate is $\dot M_a = \dot M_{\rm in} + 
 \dot M_{\rm out}$. These mass-flux rates are actually regarded as those 
 in the disk region, since $\dot M_{\rm jet}$ is a few orders of magnitude 
 smaller than $\dot M_{\rm in}, \dot M_{\rm out}$, and $\dot M_{\rm a}$ 
 in figure 7(e).  
 Similarly,  $\dot M_{\rm in}(\zeta)$ and $\dot M_{\rm out}(\zeta)$ are defined as
 
 \begin{equation}
       \dot M_{\rm in}(\zeta) = 4\pi {\rm cos} \zeta \int_{R_*}^{R_{\rm max}} 
       {r\;\rho\; {\rm min}\;(w,0)} dr, 
 \end{equation}
 \noindent and
 \begin{equation}
       \dot M_{\rm out}(\zeta) = 4\pi {\rm cos} \zeta \int_{R_*}^{R_{\rm max}} 
       {r\;\rho\; {\rm max}\;(w,0)} dr.
 \end{equation}

  \citet{Abnara2001} suggests that low-viscosity accretion flows 
  around black holes consist of two zones: an inner advection-dominated 
  zone, in which the net mass-inflow rate, $\dot M_{\rm in}$, is small, 
  and an outer convection-dominated zone, in which $\dot M_{\rm in}$ 
  increases with increasing radii.
  From figures 6 and 7, we can see that the accretion flow apparently 
  consists of two such zones.
  The transition radius, $r_{\rm tr}$, between the zones is $\sim 
  10R_*(=20r_{\rm g})$, which is larger than $10r_{\rm g}$ in 
  Stone and Pringle (2001), but smaller than 30--50$r_{\rm g}$ 
  in Abramowicz et al. (2001).
  At $r \lesssim r_{\rm tr}$, convection is absent and the mass-inflow rate is 
  very small.  On the other hand, at $r \ge r_{\rm tr}$ the flow is
  convectively turbulent and accretes slowly.
   
  From figure 7, we have rough approximations of $\rho\; \propto \;r$,
  $P_{\rm T} \;\propto \;r^{-0.85}$, $v_{\varphi}\; \propto \;r^{-1/2}$, and 
  $\dot M_{\rm in} \;\propto \;r$.  The angle-averaged variables are averaged 
  over twenty mesh points between $0 \le \zeta \le 12^{\circ}$.  
  However, it should be noticed that some variables, such as the entropy, 
  may be considerably dependent on $\zeta$, as is found in figure 7g, 
  which shows $S(\zeta \sim 0) \propto r^{-1.6}$ at $r/R_* \gtrsim 12$ but
  $S(\zeta= 42^{\circ}) \sim $ constant at $r/R_* \gtrsim 40$. 
  The entropies at $\zeta=0$ and $42^{\circ}$ sharply 
  increase at $r/R_* \lesssim$ 12 and 40, respectively, which lie almost in 
  the disk surface, because of  $S \propto T^3/\rho$ in the radiation-dominated   region.
  The entropy profiles within the disk show  equi-contours vertical
  to the equator and considerable radial gradients near to the disk mid-plane, 
  but radial equi-contours in the upper disk. 
  
  Other radial profiles are compared with
  those of (1) $\rho \;\propto \;r^{-1/2}$, $P_{\rm t} \;\propto \;r^{-3/2}$, 
  $v \;\propto \;r^{-1/2}$, and $\dot M_{\rm in}\; \propto \;r$ for 
  a self-similar solution (\cite{Igumen1999}; \cite{Narayan2000}),
  (2) $\rho \;\propto \;r^{-1/2}$, $P_{\rm t}\; \propto \;r^{-3/2}$, 
  $v = 0$,  $v_{\rm \varphi}\;\propto \;r^{-1/2}$, and $\dot M_{\rm in}$ =0 
  for a non-accretion convective envelope  solution (\cite{Narayan2000}; 
  \cite{Quataert2000}), and (3) $\rho \;\propto \;r^{0}$, $P_{\rm t} \;
  \propto \;r^{-1}$, $v\; \propto \;r^{-1}$, and $\dot M_{\rm in} \;\propto 
  \; r$ for hydrodynamical simulations with $\nu = 10^{-2}\rho$ 
  (\cite{Stone1999}).  
  The large differences of the profiles between us and them are the density
   profile with $\rho \propto r$ in model BH-1.
  However, it is natural that the very luminous disk like model BH-1 
  would have such density inversion profile, 
  because the initial Shakura--Sunyaev disk has $\rho \propto r^{3/2}$ and 
  $P_{\rm t} \propto T^4 \propto r^{-3/2}$ in the inner region of the disk, 
  where the radiation-pressure and elctron scattering are dominant 
  (Shakura, Sunyaev 1973).
  However, the self-similar solutions and other simulations are considered 
  under a negligible radiation-pressure condition.

  In figure 6, we remark on the existence of an  "accretion zone"
  just below the disk boundary, but above the top of the convective zones. 
  This accretion zone is also found in Eggum et al. (1985, 1988).  
  At $r > r_{\rm tr}$, roughly half the mass at any time at any radius will 
  be flowing in and flowing out, respectively,
  and the mass-inflow rate, $\dot M_{\rm in}$, balances the mass-outflow rate,
  $\dot M_{\rm out}$.  Near to the transition region, by way of the convective 
  zones and the accretion zone, matter is  accreted towards the 
  equatorial plane.  
  The accreting matter, which is carried to the transition region 
  by convection, partly diverts into the high-velocity jets and partly flows 
  into the inner advection-dominated zone.
  Thus, the  mass-flow rate swallowed into the black hole is as very 
 small as $\sim 5 \times 10^{14}$  g  s$^{-1}$, because the densities
 near the inner boundary are as very small as  $\sim 10^{-14}$--$10^{-26}$
 g cm$^{-3}$, although the velocities are as very large as $-0.3 c$ to $-0.6 c$. On the other hand, the mass-loss rate of the jets at the outermost boundary
 is  $\sim 4 \times  10^{18}$ g  s$^{-1}$, which is one order of 
 magnitude smaller than the input accretion rate, $\dot M_*$, whereas
 the mass-outflow rate, $\dot M_{\rm out}$, through the 
 outermost disk boundary is as large as the input accretion rate.
 We notice that the radial mass-outflow rate in the jets region
 increases with radii, roughly as $\dot M_{\rm jet} \propto r$ in figure 7e, 
 analogously with other mass-flow rates $\dot M_{\rm in}(r)$ and $\dot 
 M_{\rm out}(r)$, and furthermore that the $\zeta$-direction net mass-outflow 
 rate $\dot M_{\rm p}$ in figure 7h changes from its negative value 
 to a positive value of $\sim 10^{18}$ g s$^{-1}$  at $\zeta \sim 
 56^{\circ}$, where the angle position 
 corresponds to the boundary between the disk and the jets region. 
 After crossing the boundary, $\dot M_{\rm p}$ decreases abruptly
 because the outflow gas from the disk surface is strongly bent 
 toward the radial direction  due to the dominant radial radiation-pressure 
 force in the high-velocity region. 

 This shows that a considerable disk wind is generated in the upper disk 
 and is incorporated into the high-velocity flow. Therefore, a part of 
  the convective flow  escapes from the whole system as the disk 
  wind and the other may always remain in convective circulations 
  through the disk.  
 Thus far, we expect that the  mass-loss rate of the jets
 at great large radii will become much larger than the $ 4 \times  10^{18}$ g  
 s$^{-1}$  calculated here.

 \subsection{Model BH-2 with $\alpha=0.1$}
 
 \begin{figure}
 \begin{center}
 \FigureFile(85mm,80mm){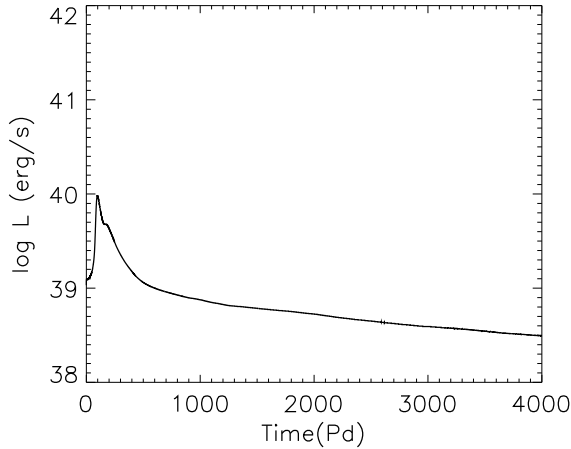}
 \end{center}
 \caption{Time evolution of the luminosity $L$ for model BH-2. }
  \label{fig:fig8}
 \end{figure}
 
 The shape of the initial disk is very similar to that of  model BH-1,
 but the densities and the temperatures on the disk mid-plane  
 in model BH-2 are  more than one order of magnitude and 
 by a few factors larger than those in model BH-1, respectively.
 Figure \ref{fig:fig8} shows the time evolution of the luminosity 
 for model BH-2.
 The luminosity,  $ 2.5 \times 10^{38}$  erg  s$^{-1}$,  at the
 final phase, $t=4365 P_{\rm d}$, is one order of magnitude smaller 
 than that in model BH-1, and $L/L_{\rm E}$ is $\sim 0.16$.

 This model shows  very different time evolutions from model BH-1.
 Mass ejection from the disk surface first begins  at 
 the innermost region of the disk, and subsequently occurs at the outer 
 part of the disk.
 Although the ejected gas is blown off partly through the outer boundary,
 it accumulates in the outer part of the disk  with increasing time. 
 Finally, the disk becomes fatter at the outer part than the initial disk.  
 The obtained accretion disk is geometrically thick as $H/r \sim 
 0.2$, and has an opening angle of $\sim 10^{\circ}$ to the equatorial plane.
 
  \begin{figure}
 \begin{center}
 \FigureFile(85mm,155mm){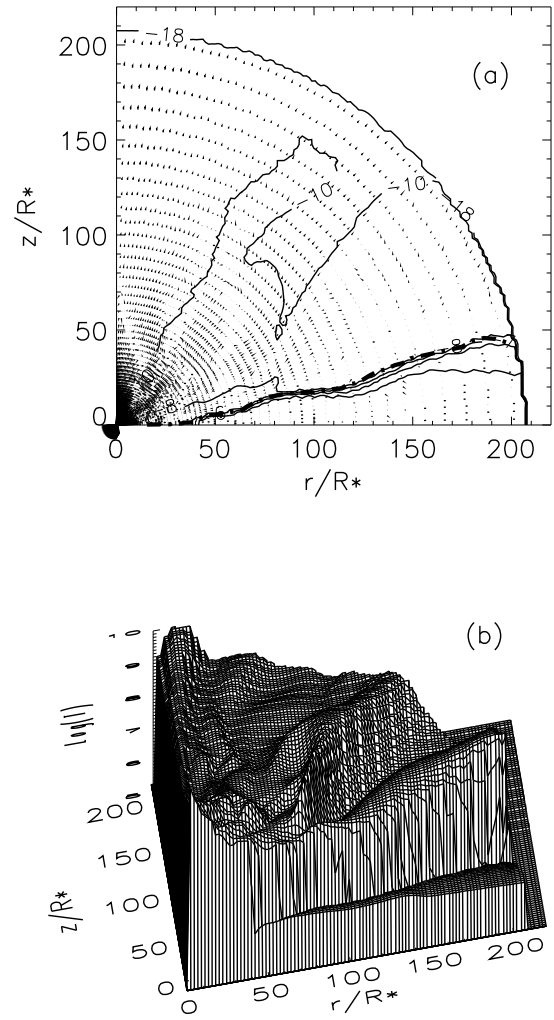}
 \end{center}
 \caption{Velocity vectors and contours of the density $\rho$
 (g s$^{-3}$) in logarithmic scale (a) and bird's-eye
 views of the gas temperature $T$ (K) (b) on the meridional plane
 at $ t= 4365 P_{\rm d}$ for model BH-2, where
 the thick dot-dashed line in (a) shows the disk boundary between
 the disk and the surrounding atmosphere, and the flow vectors
  are denoted in the same units as in figure 3a.
 }
 \label{fig:fig9}
 \end{figure}
 
  Figure \ref{fig:fig9} shows the density contours with 
  velocity vectors (a) and
 bird's-eye views of the temperature contours (b) on the meridional
 plane at the final phase.  
 The thick dot-dashed line in figure 9a shows the disk boundary 
  between the disk and the surounding atmosphere.
 In this figure we never find any relativistic jets along 
 the rotational axis as is found in model BH-1.
 The temperatures and densities at the atmospheric region around
 the disk are as hot as $\sim 10^8$--$10^{10}$ K and as low as 
 $\sim 10^{-9}$--$10^{-12}$ g cm$^{-3}$, respectively, but the 
 outflow velocities are not very large in model BH-2, where the maximum 
 velocity is $\sim$ 0.07 $c$ near the rotational axis.
 Large gradients of the density and the temperature across
 the disk boundary are also found in figures 9a and b.
 
\begin{figure}
 \begin{center}
 \FigureFile(85mm,155mm){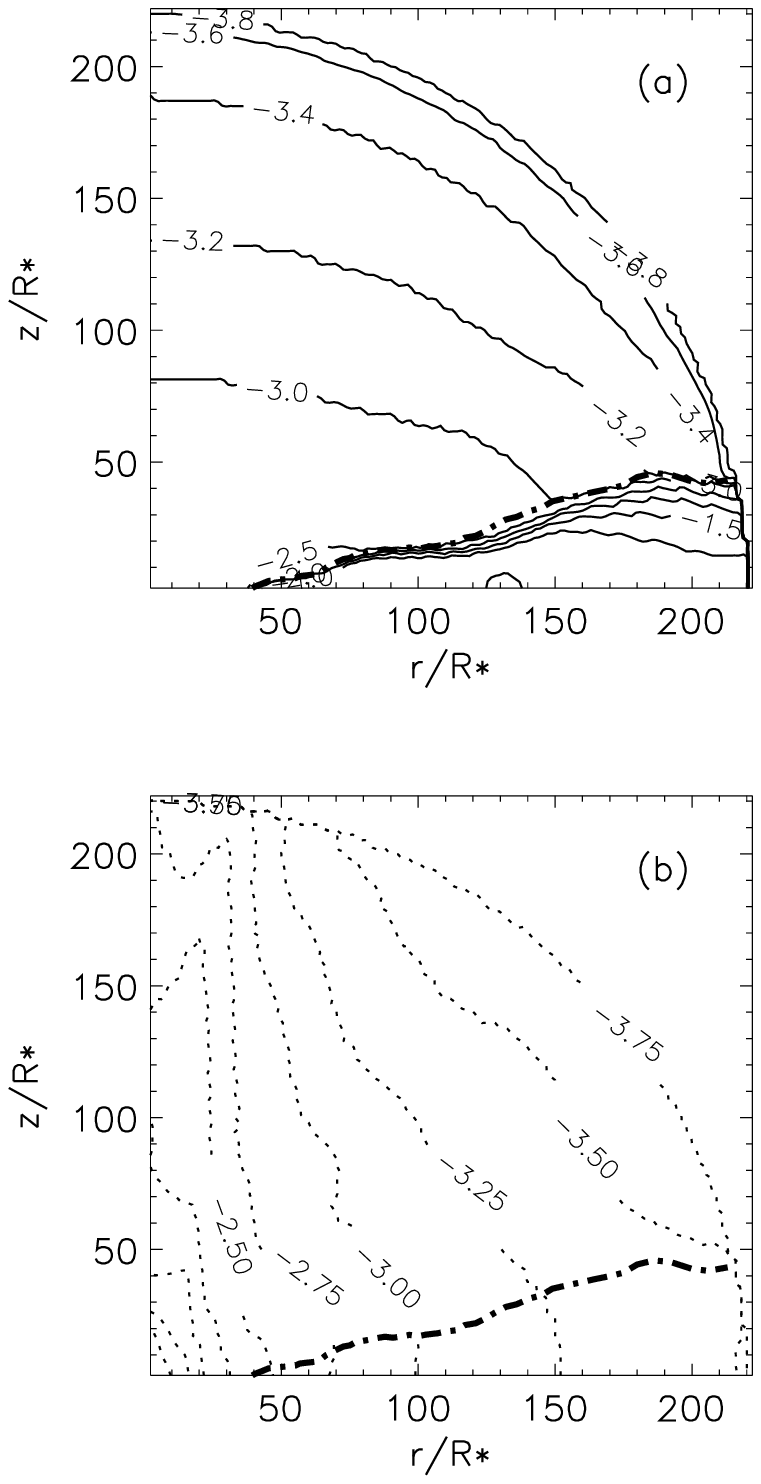}%
 \end{center}
 \caption{ Contours of the radiation energy density (a) and
 the angular-velocity $\Omega$ (b) on the meidional plane at $t=
  4365 P_{\rm d}$ for model BH-2 with the same units as in figure 4.
  }\label{fig:fig10}
 \end{figure}

 Figure \ref{fig:fig10} shows  contours of the radiation energy 
 density $E_0$ (a) and the 
 angular velocity $\Omega$ (b) on the meridional plane at
 $t=4365 P_{\rm d}$ with the same units as in figure 4.
 The disk is nearly Keplerian throughout its whole shape,
  but the atmospheric region above the disk shows 
  non-Keplerian angular velocities.
 There also appear  convective cells at $15 \lesssim r/R_* \lesssim 50$.
 The densities and temperatures at $r/R_* \lesssim 10$ on the disk 
 mid-plane become more rarefied and much hotter than those in the initial 
 disk, respectively,  and a spherical high-temperature region is formed 
 around here.
  The densities near the inner boundary are much larger, 
  $ \sim 10^{-9}$ -- $10^{-10}$ g cm$^{-3}$, than 
  those in model BH-1. The resultant mass-inflow rate swallowed 
  into the black hole is much larger, $\sim -1.4 \times 10^{16}$ g s
  $^{-1}$, than the $\sim -5\times 10^{14}$  g s$^{-1}$ in model BH-1, 
  but is negligibly small compared with the input accretion rate.
  
 The disk features obtained here are interpreted, rather, under a category of 
 the standard  disk model, though the disk is geometrically thick.
  The result of model BH-2 describes
  that the concept of relativistic jets expected for the 
 super-Eddington accretion disks around black holes 
  must be modified under the $\alpha$-model of viscosity, i.e., 
 for the case of strong kinematic viscosity, there exists no relativistic 
 and well-collimated outflow jet, even in the super-Eddington accretion disks. 
 Though the densities and temperatures at the atmosphere around the disk 
 are in the range of those required for the X-ray jets of SS 433, 
 the maximum outflow velocity  is at most $\sim 0.07c$ and the flows are not
 well collimated anywhere. 
 Therefore, we can not expect X-ray emission lines with a definite 
 Doppler shift of 0.26 $c$ observed in SS 433, and conclude that model BH-2 is 
 unfavorable for SS 433.

\subsection {X-ray Emitting Region of  Iron Lines}

 For a Maxwellian distribution of the electron velocities, the power emitted
 per unit volume due to excitations of level $n^{\prime}$ of ion $Z$ 
 in the ground state $n$ by electron collision is given by  
 \begin{equation}
       {dP\over dV} = 1.9 \times 10^{-16}T^{-1/2}\overline{\Omega}
          \left({\Delta E \over {I_{\rm H}}}\right)e^{-\Delta E/kT}
          N_{\rm e}N_{\rm Z} \;\;\; {\rm erg}\;{\rm cm}^{-3}\;{\rm s}^{-1},  
 \end{equation}
 where $\Delta E$, $I_{\rm H}$, $k$, $N_{\rm e}$, $N_{\rm Z}$, and 
 $\overline{\Omega}$  are the excitation energy between the 
 $n$ and $n^{\prime}$ levels, the ionization potential of hydrogen, 
 the Boltzmann constant, the number density of electrons, the number 
 density of ion $Z$, and the effective average of collision strength 
 $\Omega$, respectively (\cite{Blume1974}).
 The red- and blue-shifted highly ionized iron lines, FeXXV K$_{\alpha}$ 
 and Fe XXVI K$_{\alpha}$, are clearly seen in all of the observed X-ray
 spectra of SS 433 (\cite{Kotan1996}). 
 Typically, focusing on the Fe XXV K$_{\alpha}$ line, we calculated the 
 total power emitted by the iron lines through
 the optically thin region.
 
 \begin{figure}
 \begin{center}
 \FigureFile(85mm,80mm){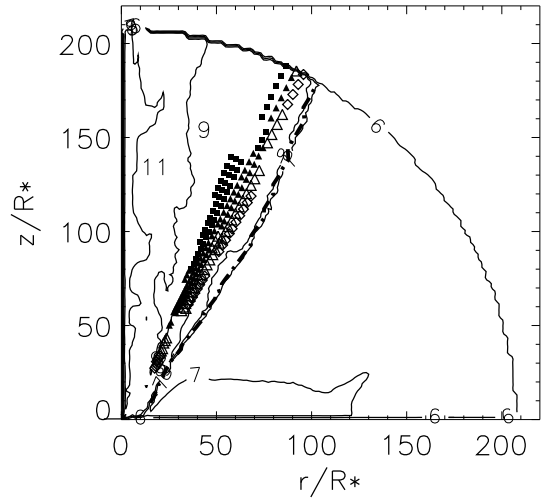}
 \end{center}
  \caption{Distribution of mesh points contributed mostly to the total X-ray
  emission of iron lines at $t$ = 3577$P_{\rm d}$ 
  for model BH-1, where the disk boundary (thick dot-dashed line)
  and the temperature contours of log $T$ = 7, 8, 9, and 11 (lines) are 
  also plotted. 
  The diamonds, triangles, filled triangles, and filled squares 
  denote the mesh points with  different flow velocities of $\sim$ 0.10 $c$, 
  0.13 $c$, 0.17 $c$, and 0.20 $c$, respectively. 
  Only mesh points with emissivity of 
  more than a hundredth of the maximum emissivity in the optically 
  thin region are plotted here.}\label{fig:fig11}
  \end{figure}
 
 Figure \ref{fig:fig11} shows the mesh points which contributed most 
 to the total X-ray power of the iron lines at 
 $t = 3577 P_{\rm d}$ for model BH-1,
 where only  mesh points with emissivity of more than a hundredth 
 of the maximum 
 emissivity in the optically thin region are indicated  by the 
 triangles, squares, filled triangles, and filled squares, 
 which show different flow velocities of $\sim$ 0.10 $c$, 0.13 $c$, 
 0.17 $c$, and 0.20 $c$ at the mesh points, respectively.
 The influences of photoionization and photoexcitation by the intense
  radiation field in the innermost region were not taken account of here.
 If these effects are considerable, the results in figure 11
 may be invalid. In this respect, we mention that \citet{Brink2000} showed
 the effect of photoionization on the iron-line flux ratio to be 
 relatively small, using their jets model of SS 433 and assuming that the 
 emission at the base of the jets is given by a black body, where
 the temperatures and densities ($4\times 10^8$ K and $10^{-9}$--$10^{-10}$ g cm$^{-3}$, respectively)  used in their model are of the same orders
 as those in the marked region in figure 11.   
 
 We notice that the emitting region of the iron lines in figure 11
  is confined into a narrow region at $63^{\circ} \lesssim \zeta 
  \lesssim 68^{\circ}$ just outside the  disk boundary, 
 and that the high-velocity gas in the funnel region would never contribute 
 to the X-ray line emission, because the densities in the  
 region are more than ten orders of magnitude smaller than those 
 in the narrow region, regardless of the temperatures in the funnel 
 region. 
 The accretion disk, itself,  does not contribute to the 
 X-ray line emission, because the temperatures are as low as 
 $\sim 10^6$--$10^7$ K and its emissivity decreases exponentially as 
 $e^{-\Delta E/kT}$ in spite of the high densities in the disk.
 Therefore, we consider that the unique velocity, 0.26 $c$, of SS 433 would 
 be attributed to the effective velocity in the above emitting region of 
 the iron lines.


\section{Discussion}
\subsection{Comparison with Other Results}

 Our present results  are  compared with those
 of a super-Eddington black hole model
 by Eggum et al. (1985, 1988), who considered
 a black hole  of $M_*=3M_{\solar}$ and $\dot M_*=4\dot M_{\rm E}$
 and used the flux-limited diffusion approximation for the radiation
 transport. The black hole mass is by a factor 3 smaller than ours, but
 the input accretion rate is the same as ours.
 They used a constant kinematic viscosity, $\nu = 1.5\times 10^{15}$ cm$^2$ 
 s$^{-1}$, instead of the usual $\alpha$-model for viscosity.
 
 There exist some qualitatively important and common results 
 between our model BH-1 and their model, such as the relativistic funnel jets, 
 the geometrically thick disk, the accretion zone, and the remarkable 
 convective phenomena in the disk.  
 In their results, the flows between the disk and the rotational axis are 
 accelerated to relativistic velocities due to the radiation-pressure force.
 The collimation angle of the jets to the rotational axis is also as  
 large as $\sim 30^{\circ}$ in model BH-1.
 The mass-outflow rate obtained by Eggum et al. (1985, 1988)
 is three orders of magnitude smaller than the input accretion rate,
  $\dot M_*$, and accordingly the estimated $\dot M_{\rm loss}$ of  SS 433.
 On the other hand, the mass-outflow rate from the high-velocity
  jets in model BH-1 is much larger than that in Eggum et al. although it 
  is one order of magnitude smaller than the estimated value in SS 433.
  Most of the accreting matter in Eggum et al. (1985, 1988) is swallowed into
 the central black hole through the inner boundary,
 but the mass-flow rate into the 
 black hole in our model is negligibly small compared with the input
 accretion rate. 
 The differences in the numerical results may be attributed
 to the high kinematic-viscosity, $\nu$, of $1.5 \times 10^{15}$ cm$^2$ 
 s$^{-1}$ in Eggum et al. (1985, 1988), which is more than two 
  orders of magnitude larger than those at $r/R_* \lesssim 20$ and 
  at $r/R_* \gtrsim 50$, respectively, on the disk mid-plane in model BH-1.

\subsection {Convection-Dominated Accretion Flows with Small $\alpha$}

 As an alternative model to ADAFs models, since \citet{Narayan1994},
 \citet{Bland1999} proposed advection-dominated inflow-outflow 
 solutions (ADIOS) which include a powerful wind that carries away mass, 
 angular momentum, and energy from the accreting gas. 
 A typical case with appropriate parameters shows that only a small 
 fraction of the mass supplied will reach the black hole, accompanying 
 a disk wind. 
 In relation to these ADAFs, \citet{Narayan2000}, Quataert and Gruzinov(2000),
 and \citet{Abnara2001} developed the convection-dominated accretion 
 flows (CDAFs) or a 'convective envelope solution'. 
 The CDAFs models assume that the convective energy transport
 is primarily directed radially outward through the disk, 
 but with no outflows, whereas in the ADIOS models the convective 
 energy transport has a large $\zeta$-direction component toward 
 the disk surface at high latitudes, which will drive strong outflows.
  
 The initial Shakura--Sunyaev disk for model BH-1 is originally unstable
 to convection, since the entropy, $S \propto T^3/\rho$, in the 
 radiation-pressure dominant inner disk falls abruptly with increasing $z$. 
 Convection tends to establish an isentropic structure along the $Z$-axis, 
 which results in equi-contours of entropy vertical to the equator near to the 
 disk mid-plane in model BH-1. 
 The convective flows accompanied by a large mass-outfow from the disk 
 surface  may be relevant to the ADIOS models.  
 On the other hand, large negative radial gradients of the entropy near to 
 the mid-plane in model BH-1 also induce large convective energy 
 transport directed radially outward through the disk, as can be seen in 
 figures 7f and g.
 As a result, the accretion flows in model BH-1 may be dominated by a 
 ADIOS model and also a CDAFs model.
 
 The convective envelope solution, which is special one of the CDAFs,
 was  addressed by \citet{Narayan2000} and Quataert and Grunizov (2000).
 The convective envelope solution expresses that 
 $v \sim 0$ and the viscous dissipation rate, $Q^+$, is $\sim 0$ because  the
 net shear stress via viscosity and via convection vanishes. 
 This leads to no advection entropy,
 and thus the divergence of the convective energy flux vanishes from 
 the energy equation, that is, $F_{\rm c}(r) \propto r^{-2}$.
 The relation $F_{\rm c} \propto r^{-2}$ is 
 also derived from the definition $F_{\rm c} = -3\alpha (c_{\rm s}^2/
 \Omega_{\rm K}) \rho T dS/dr$ (\cite{Narayan2000})
  and using the radial profiles of $\rho \propto r$ and $P_{\rm t} \propto
  r^{-0.85}$ in figure 7, where $c_{\rm s}$ is the sound speed. 
 Figure 7f shows roughly $L_{\rm c} \sim$ constant in the outer region; 
 accordingly,  $F_{\rm c} \propto r^{-2}$ if $F_{\rm c}$ is independent 
 of $\zeta$.  Even for net zero stress, the negative radial gradient of
 the entropy would be maintained if the high-entropy gas is lost in a 
 considerable disk wind, as is found in model BH-1.
 Accordingly,  model BH-1 may possess a circumstance near the convective 
 envelope solution, although the mean radial velocity is not exactly zero.
 In this respect, further knowledge of the actual mechanism
 of the angular momentum transport by convection is required.  
 
  Abramowicz and Igumenshchev (2001) suggest that CDAFs have
  a significant outward energy flux carried by convection, with a 
  luminosity of $L_{\rm c} = \epsilon_{\rm c} \dot M c^2$, 
  where the efficiency, $\epsilon_{\rm c}$, is $\sim 3\times 
  10^{-3}$--$10^{-2}$,  independently of the accretion rate, and  
  the radiative output comes mostly from the convective part. 
  In model BH-1, the convective luminosity, $L_{\rm c}$, near the outer 
  boundary is $\sim 5\times 10^{37}$ erg s$^{-1}$, which is far smaller than
  the radiative luminosity, $L (\sim 1.6\times 10^{39}$ at $t$ = 
  3577 $P_{\rm d}$ in model BH-1);  this corresponds 
  to $\epsilon_{\rm c} \sim 10^{-3}$, which agrees with that by 
  Abramowicz and Igumenshchev (2001). 
   However,  the luminosity, $L$,  comes mostly from
   the innermost hot region at $r \sim r_{\rm tr}$ through the optically thin 
   high-velocity region, instead of the convective-dominated disk. 
   Indeed, $\dot M_aGM/r_{\rm tr}$  gives $2\times 10^{39}$ erg s 
   $^{-1}$, which corresponds well to $L$, 
    roughly using  $\dot M_a(r_{\rm tr}) \sim 10^{20}$ g s$^{-1}$ and 
    $r_{\rm tr} \sim 20r_{\rm g}$ from figure 7.  
   
  Finally, we emphasize that most of the simulations and
 theoretical models so far have been done on thin accretion disks without
 radiation, and that it is not still clear whether these results are valid
 for the much thicker and radiation-pressure dominant disks. 
 Further theoretical studies on the radiation-pressure dominant ADAFs and 
 CDAFs are required.

 \subsection{Relativistic Jets, Collimation, and Mass-Outflow Rate}
  
  Observations of SS 433 show many emission lines of heavy elements, such
  as Fe and Ni, which denote the unique red and blue Doppler shifts of 0.26 $c$
  in the X-ray region. 
  The collimation angle of SS 433 jets seems to be as small 
  as 0.1 radian (several degrees) (\cite{Margo1984}).
 Any models for SS 433 must explain the characteristics of these emission 
 lines. The unique velocity, 0.26 $c$, may be reasonably explained in terms of 
 the relativistic velocities at a confined  region just outside of the upper 
 disk boundary, which lies between the funnel wall and the disk, as is 
 found in model BH-1.  
 The velocity  0.26 $c$ is not an inherent velocity of SS 433 like 
 an orbital velocity in the binary system, but is a result of 
 the super-Eddington accretion disk.  
 
 The collimation angle, $\sim 30^{\circ}$, of the high-velocity jets in model 
 BH-1 is rather large compared with the value of $\sim$ 0.1 radian expected for SS 433 jets.
 If the calculated high-velocity jets should be truly confined to
 a small angle of $\sim $ 0.1 radian from the rotating axis, we need a 
 collimation mechanism operating outside the present computation grids, 
 or some mechanism, such as magnetohydrodynamical collimation, which is not 
 considered here. This problem remains open, although there exist some 
 discussions that the jets of SS 433 may not be well collimated, as is 
 inferred from the line profiles, and that the 0.1 radian jet may not be
 consistent with the X-ray observations of SS 433 (Eggum et al. 1988).
 
  Hydrodynamical modeling of the jets combined with recent X-ray
 observations reveals temperatures of 6--8$\times 10^8 $  K, particle 
 densities of between $5\times 10^{11}$--$5\times 10^{13}$ cm$^{-3}$ at 
 the base of the jets, and the length of the X-ray jets being 
 $ \gtrsim 10^{10}$ cm (\cite{Brink1991}; \cite{Brink1993}; \cite{Kotan1996}).
  These temperatures and densities required for the X-ray jets
 qualitatively agree with those at the high-velocity region 
 between the funnel wall and the disk boundary in model BH-1. 
 The X-ray emitting region observed in SS 433 would be in 
 the range of $ r = 10^{10}$--$10^{12}$ cm (\cite{Kotan1996}), which 
 far exceeds the present computational domain.
 The high-velocity jets region in model BH-1 would probably extend up to 
 $ r  \sim 10^{12} \;$cm. 
 Additionally, from the characteristics of the results given in figure 3, 
 we expect that such an extended high-velocity region would have 
 temperatures and densities of $\sim 10^8$ K and $\sim 10^{-12}$--$10^{-10}$ g cm$^{-3}$.
 Although the mass-loss rate ($4 \times 10^{18}$ g s$^{-1}$) of the jets 
 in model BH-1 is one order of magnitude smaller than the mass-loss rate 
 ( $\ge 3 \times 10^{19}$ g  s$^{-1}$) estimated in SS 433, we speculate that
 the actual calculated mass-loss rate at large radii may become comparable
  to the observed one, because the jets mass-flux would increase due to the 
  wind gas from the disk at large radii.
 In such a case, we may have a problem that the jets gas may slow down below 
 0.26 $c$ if the enhanced region of the jets mass-flux is too far from the 
 present computational domain. 
 Alternatively, the super-Eddington models with a much larger
 input accretion rate, $\dot M_*$, than the present value, $4\; \dot M_{\rm
 E}$, may be responsible for a mass-loss rate comparable to $\sim 10^{20}$ g s$^{-1}$.

\subsection{ Central Object of SS 433 }

  Whether the central object of SS 433 is a black hole or a neutron
  star remains unsolved even at present, because we have not yet
  obtained  any decisive evidence of a central source mass.
 From some observational and theoretical estimates of a massive mass and 
 its too high energetics
 of SS 433, many astrophysicists seem to favor a black-hole
 hypothesis of SS 433 (\cite{Leibo1984}; \cite{Fabri1990}
 ;  \cite{Chere1993}; \cite{Fukue1998}; \cite{Hirai2001}),
 while some people suggest that the compact object is a
 neutron star, from the viewpoint of a theoretical model and
observations of the He {\sc ii} line (\cite{Begel1984}; \cite{D'Odo1991}; 
\cite{Zwitt1993}).

 In this respect, our present results are also compared with those of 
 a neutron-star model by \citet{Okuda2000} with 
 $M_* = 1.4 M_{\solar}$,  $R_* = 10^6 \;{\rm cm}$, and a 
 mass accretion rate ($\dot M_*$) of $10^{20}\; {\rm g}\; {\rm s}^{-1}$, which 
 corresponds to $\sim 100 \dot M_{\rm E}$ for a neutron star. 
 The viscosity parameter, $\alpha = 10^{-3}$, used is the same as that in 
 model BH-1.  
 The characteristic features in model BH-1 are also found in this 
 neutron-star model, where there appear a high-velocity jet region 
 with a collimation angle of $\sim 10^{\circ}$ to the rotational axis, 
 the existence of a confined X-ray emitting region of iron lines in the 
 high-velocity region, a geometrical thick disk with an opening angle 
 of $80^{\circ}$ to the equatorial plane, a mass-loss rate far less than 
 $ 10^{20}\; {\rm g} \;{\rm s}^{-1}$, and  convective motions in the 
 inner disk. Regarding the collimation angle, the neutron-star
 model with a smaller angle $\sim 10^{\circ}$ may be favorable for
 SS 433.  However, except for this point, there is no intrinsic or large
 quantitative difference between the neutron-star model and the black-hole 
 model; that is, we could not find any decisive proof for these objects 
 as a candidate of SS 433. 

\subsection{Assumption of the Flux-Limited Diffusion Approximation}

The radiation transport was treated here as an approximation of flux-limited
 diffusion. The flux limiter, $\lambda$, and the Eddington factor, $f_{\rm E}$,
  used in the approximation are given from the empirical formulas fitted to
 some stellar atmosphere models (\cite{Kley1989}).  
 Recently, \citet{Turner2001} addressed that the flux-limited diffusion
 approximation (FLD) is less accurate when the flux has a component 
 perpendicular to the gradient in radiation energy density, and in optically 
 thin regions when the radiation field depends strongly on the angle.  
 The calculations considered in our case  typically result in two optical 
 regions, which are the optically thick disk region and the optically thin 
 high-velocity region. The flux limiter, $\lambda$, is almost 1/3 in the disk 
 and, contrarily, very small everywhere in the high-velocity region. 
 In an optically thick disk, FLD is sufficient. 
 On the other hand, in the optically thin high-velocity region, the radiation 
 fields are highly anisotropic and FLD may be less accurate.  
 The radiation fields in the high-velocity region originate in radiation 
 from the  central hot region and the outer disk surface. 
 However, the contours of radiation energy density are almost concentric 
 around the central object. 
 Actually, the radial radiation flux, which is attributed to the central hot 
 region, is one order of manitude larger than the radiation flux normal 
 to the disk surface.
 As a result, we consider that FLD in our code would not result in any serious
 influences on the whole dynamics of the flows and the total luminosity;
 still, this problem should be checked with a more accurate method in future 
 improvements of the numerical code.
 
\section{Conclusions}

 We examined  highly super-Eddington black-hole models for SS 433, based on
 two-dimensional hydrodynamical calculations coupled with  
 radiation transport. A model with a  small viscosity parameter, $\alpha =
  10^{-3}$, shows that a geometrically and optically thick convection-dominated
  disk with a large opening angle of $\sim 60^{\circ}$ to the equatorial plane 
  and  rarefied, very hot, and optically thin high-velocity jets region 
  around the disk are formed. 
  The thick accretion flow near to the equatorial plane consists of 
  two different zones: an inner 
  advection-dominated zone, in which the net mass-inflow rate, 
  $\dot M_{\rm in}$, is very small, and an outer convection-dominated 
  zone, in which $\dot M_{\rm in}$ increases with increasing radii.
  The high-velocity region along the rotating axis is divided into 
  two characteristic regions 
  by the funnel wall, a barrier where the effective potential
  due to the gravitational potential and the centrifugal one vanishes.
  A confined region in the high-velocity 
  jet region just outside of the photospheric disk boundary 
  would be responsible for the observed X-ray iron emission lines with 
  a Doppler shift of 0.26 $c$.
  However, from this model, we can not obtain a small collimation 
  degree of $\sim$ 0.1 radian of the jets and a sufficient mass-outflow rate 
  of the jets comparable to $\sim 10^{20}$ g s$^{-1}$, as is expected 
  for SS 433.  These problems still remain  open.
  On the other hand, from a model with a large $\alpha$ = 0.1, we find 
  a geometrically and optically thick quasi-Keplerian disk with an opening
  angle of $\sim 10^{\circ}$ to the eauatorial plane, whereas the disk is far 
  thinner than the disk with $\alpha = 10^{-3}$. 
  This model may be unfavorable to SS 433  
  because we never find  relativistic jets with $\sim$ 0.2 $c$ here.
  Further investigations of the super-Eddington models with other model
  parameters and over a wider range of the computational domain 
  ( $r \sim 10^{12}$ cm) may give some key to the open problems.

 \bigskip
 \bigskip
 The author gratefully thanks Dr. D. Molteni for his useful
 discussions of the numerical method and the results, and also the referee for 
 many useful comments concerning the advection-dominated and the 
 convection-dominated accretion flows.
 Numerical computations were carried out at the Information Processing Center 
 of Hokkaido University of Education.

\end{document}